\documentclass[12pt]{iopart}

\usepackage{graphicx}
\usepackage{amssymb}
\usepackage{iopams}

\newcommand{\dfrac}[2]{\case{\displaystyle #1}{\displaystyle #2}}

\newcommand{\email}[1]{\ead{#1}}
\newcommand{\affiliation}[1]{\address{#1}}
\newcommand{\acknowledgments}{\ack}

\newcommand{\setR}{\mathbb{R}}

\newcommand{\uc}{\mathrm{c}}
\newcommand{\ub}{\mathrm{b}}

\newcommand{\zero}{{_0}}
\newcommand{\one}{{_1}}
\newcommand{\two}{{_2}}

\newcommand{\dm}{\mathrm{dm}}
\newcommand{\sr}{\mathrm{sr}}
\newcommand{\wig}{\mathrm{wig}}
\newcommand{\Lcb}{\bar{\mathcal{L}}}

\newcommand{\calO}{\mathcal{O}}
\newcommand{\Mc}{M_\uc}
\newcommand{\OmegaL}{\Omega_\Lambda}
\newcommand{\OmegaCDM}{\Omega_\dm}
\newcommand{\OmegaB}{\Omega_\ub}
\newcommand{\zre}{z_{\mathrm{re}}}
\newcommand{\Ps}{P_\mathrm{scalar}}

\newcommand{\CAMB}{\texttt{CAMB} }
\newcommand{\COSMOMC}{\texttt{COSMOMC} }
\def\setC{\mathbb{C}}

\def\setR{\mathbb{R}}

\newcommand{\ie}{\textsl{i.e.}}
\newcommand{\eg}{\textsl{e.g.}}
\newcommand{\GReCO}{${\cal G}\setR\varepsilon\setC{\cal O}$}

\begin{document}

\title{Exploring the Superimposed Oscillations Parameter Space}

\author{J\'er\^ome Martin} 
\email{jmartin@iap.fr}
\affiliation{Institut d'Astrophysique de Paris, \GReCO --FRE 2435, 
98bis boulevard Arago, 75014 Paris, France} 

\author{Christophe Ringeval} \email{c.ringeval@imperial.ac.uk}
\affiliation{Blackett Laboratory, Imperial College London, Prince
Consort Road, London SW7~2AZ, United Kingdom}

\date{\today} 

\begin{abstract}

The space of parameters characterizing an inflationary primordial
power spectrum with small superimposed oscillations is explored
using Monte Carlo methods. The most interesting region
corresponding to high frequency oscillations is included in the
analysis. The oscillations originate from some new physics taking
place at the beginning of the inflationary phase and characterized by
the new energy scale $\Mc$. It is found that the standard slow-roll
model remains the most probable one given the first year WMAP data. At
the same time, the oscillatory models better fit the data on average,
which is consistent with previous works on the subject. This is
typical of a situation where volume effects in the parameter space
play a significant role. Then, we find the amplitude of the
oscillations to be less than $22\%$ of the mean amplitude and the new
scale $\Mc$ to be such that $H/\Mc<6.6 \times 10^{-4}$ at 1$\sigma $
level, where $H$ is the scale of inflation.

\end{abstract}

\maketitle

\section{Introduction}

The possibility that the Cosmic Microwave Background (CMB) data
contain unexpected features which would be signatures of non standard
physics has recently been studied in
Refs.~\cite{Martin:2003sg,Martin:2004iv}. These features generally
consist of superimposed oscillations in the CMB multipole moments
$C_{\ell }$. They could originate either from trans-Planckian effects
taking place in the early universe~\cite{Martin:2000xs,
Brandenberger:2000wr, Niemeyer:2000eh, Kempf:2000ac,
Kempf:2001fa,Easther:2001fi, Lemoine:2001ar, Hui:2001ce,
Easther:2001fz, Niemeyer:2001qe, Lizzi:2002ib,Danielsson:2002kx,
Hassan:2002qk, Easther:2002xe, Niemeyer:2002kh,Kaloper:2002cs,
Bergstrom:2002yd,Goldstein:2003ut,
Alberghi:2003am,Armendariz-Picon:2003gd,Martin:2003kp,Collins:2003zv,
Elgaroy:2003gq,Kaloper:2003nv,Collins:2003mj,
Okamoto:2003wk,Ashoorioon:2004vm,Brandenberger:2004kx,
Danielsson:2004xw,Greene:2004np} or from other phenomena, for instance
of the type of those investigated in
Refs.~\cite{Barriga:2000nk,Wang:2002hf,Burgess:2002ub,Martin:2003sf,
Martin:2003bp,Hunt:2004vt}.

\par 

In Refs.~\cite{Martin:2003sg,Martin:2004iv}, it has been demonstrated
that the presence of superimposed oscillations can improve the fit to
the first year Wilkinson Microwave Anisotropies Probe (WMAP)
data~\cite{Peiris:2003ff, Bennett:2003bz, Hinshaw:2003ex,
Verde:2003ey, Kogut:2003et}, see also Refs.~\cite{Kogo:2003yb,
Shafieloo:2003gf}. The corresponding drop in the $\chi ^2$ was found
to be $\Delta \chi ^2\simeq 10$. Then, based on the so-called F-test,
it has been argued that this drop is statistically
significant. However, it should be clear that the study carried out in
Refs.~\cite{Martin:2003sg,Martin:2004iv} has only provided a hint with
regards to the possible presence of non trivial features in the
multipole moments. In order to go further, it is necessary to explore
the parameter space in details, taking into account, of course, the
parameters characterizing the oscillations, namely the amplitude, the
frequency and the phase. This task is delicate because the numerical
computation of the multipole moments in presence of rapid oscillations
in the initial power spectrum necessitates an accurate estimation of
the CMB transfer functions and of the line of sight integrals which,
in turn, requires to boost the accuracy with which the $C_{\ell }$'s
are computed (here by means of the \CAMB
code~\cite{Lewis:1999bs}). This causes an important increase of the
computational time and instead of a few seconds, the computation of
one model now takes a few minutes. Therefore, the exploration of the
full parameter space ({\ie} the one including the cosmological
parameters characterizing the transfer functions and the primordial
parameters encoding the shape of the initial power spectra) remains
beyond the present day computational facilities capabilities, at least
in a reasonable time. For this reason, in this paper, we restrict
ourselves to the exploration of the so-called fast parameter space
only, {\ie} the space of parameters characterizing the primordial
spectrum, including of course the oscillatory parameters. These
parameters are usually called ``fast'' because restricting the
exploration to the corresponding parameter subspace requires the
calculation of the transfer functions only once, hence a huge gain in
term of computational time. However, in the present work, the correct
computation of the superimposed oscillations that are transfered from
the power spectrum to the multipole moments requires a very accurate
evaluation of the line of sight integrals. Therefore, even in the fast
parameter subspace, each determination of the $C_{\ell }$'s remains a
heavy time consuming task~\cite{Martin:2003sg,Martin:2004iv}. The main
new result of the present article is that, for the first time, the
shape of the likelihood function in this space including, and this is
a crucial point, the region containing the best fit found in
Refs.~\cite{Martin:2003sg,Martin:2004iv} is presented.

\par

The present study also allows us to gain some new insight into the
statistical significance of the superimposed oscillations. In
particular, we derive the one-dimensional marginalized probabilities
and the mean likelihoods for the oscillatory parameters. Our result
indicates that the slow-roll models are still the most probable
ones. However, the marginalized probability distribution of the
amplitude possesses a long tail originating from the fact that the
mean likelihood is peaked at values corresponding to models with
oscillations. From the above-mentioned distributions, we derive
constraints on the parameters characterizing the oscillations.

\par

Finally, some comments are in order about the previous works on the
subject~\cite{Bergstrom:2002yd, Elgaroy:2003gq,Okamoto:2003wk}. In
those references, the shape of the likelihood function in the
parameter space, including cosmological parameters, was studied and
its hedgehog shape highlighted. However, the exploration was limited
to the low frequency models in order to avoid the above-mentioned
numerical difficulties. As a consequence, the region where the fit
found in Refs.~\cite{Martin:2003sg,Martin:2004iv} lies was not
considered. As mentioned above, the exploration of this region is a
crucial ingredient for the present article and plays a role in
constraining the oscillatory parameters.

\par

This paper is organized as follows. In the next section we briefly
recall the theoretical model used to generate primordial oscillations,
together with the numerical implementations. In
Sec.~\ref{sect:constraints}, marginalized probabilities and
constraints on the fast parameters are discussed. We give our
conclusions in Sec.~\ref{sect:conc}.

\section{The fast parameter space}
\label{sect:ps}

We assume that inflation of the spatially flat
Friedman-Lema\^{\i}tre-Robertson-Walker (FLRW) space-time is driven by
a single scalar field $\phi (t)$. Then, the scalar fluctuations can be
characterized by the curvature perturbations $\zeta$ and the tensor
perturbations by a transverse and traceless tensor $h_{ij}^{\rm
(T)}$~\cite{Martin:1997zd, Martin:1999wa,
Martin:2000ak,Schwarz:2001vv}. If the trans-Planckian effects are
taken into account, the corresponding power spectra take the form of a
standard inflationary part plus some oscillatory corrections. The
standard part is parameterized by the Hubble parameter during
inflation $H$, the slow-roll parameters $\epsilon _\one$ and $\epsilon
_\two$, and the pivot scale $k_*$~\cite{Martin:1997zd, Martin:1999wa,
Martin:2000ak,Schwarz:2001vv}. The two slow-roll parameters that we
use are defined by $\epsilon _\one \equiv -\dot{H}/H^2$ and $\epsilon
_2=2(\epsilon _\one -\delta )$ with $\delta \equiv -\ddot{\phi
}/(H\dot{\phi })$. The oscillatory part is determined by three new
parameters. Two of them are the modulus $\vert x\vert $ and the phase
$\varphi $ of a complex number $x$ which characterizes the initial
conditions and the other is $\sigma _\zero=H/\Mc$, $\Mc$ being a new
energy scale. Note that $\sigma_\zero$ is evaluated at a time $\eta
_\zero$ during inflation which is {\it a priori} arbitrary but, and
this is the important point, does not depend on $k$. In the following
we will choose this time such that $k_*/a_\zero=\Mc$ where $a_\zero $
is the scale factor at time $\eta _\zero $. Explicitly, for density
perturbations, the power spectrum
reads~\cite{Martin:2003sg,Martin:2004iv}
\begin{eqnarray}
\label{pssrs2}
k^3 P_{\zeta} & = k^3 P_{\sr} \Bigg\{ 1 - 2 |x| \sigma_\zero \cos\left[
\dfrac{2 \epsilon_\one }{\sigma_\zero} \ln \left(\dfrac{k}{k_*}
\right) + \psi \right]
+ \calO(|x| \sigma_\zero \epsilon_2) \Bigg\},
\end{eqnarray}
and a similar expression for the gravitational waves.  Typically, the
amplitude of the oscillations is given by $\vert x\vert \sigma _\zero$
while the wavelength is such that $\Delta k/k=\pi \sigma
_\zero/\epsilon _\one $. As discussed in Ref.~\cite{Martin:2003sg},
the factor $x$ parameterizing the initial conditions plays an
important role in our analysis. The quantum state in which the
cosmological fluctuations are ``created'', when their wavelength
equals the new characteristic scale $\Mc $, is {\it a priori} not
known. Usually, it is assumed that this quantum state is characterized
by the choice $x=1$. In the present article, following
Refs.~\cite{Goldstein:2003ut, Armendariz-Picon:2003gd,Collins:2003zv,
Collins:2003mj}, we generalize these considerations and describe our
ignorance of the initial state by the parameter $x$ which is therefore
considered as a free parameter. From the point of view of the
statistical study presented here, this turns out to be very
important. Indeed, as mentioned in Ref.~\cite{Martin:2003sg} [see the
discussion after Eq.~(8)], this allows us to decouple the amplitude of
the oscillations from the value of the frequency which, therefore, can
now be considered as independent parameters. As a consequence, high
frequency waves no longer necessarily have a small amplitude as it is
the case if one chooses $x=1$.  Let us also notice that, despite the
fact that the parameterization of the initial conditions used is more
general, we have nevertheless restrict our considerations to the case
where $x$ is scale independent over the scales of interest.

\par

In order to compute the resulting CMB anisotropies, we use a modified
version of the \CAMB code~\cite{Lewis:1999bs} whose characteristics
and settings are detailed in
Refs.~\cite{Martin:2003sg,Martin:2004iv}. The exploration of the
parameter space is performed by using the Markov Chain Monte-Carlo
methods implemented in the \COSMOMC code~\cite{Lewis:2002ah} together
with our modified \CAMB version. As already mentioned, in order to
avoid prohibitive computational time, only the ``fast parameter
space'' is explored. Notice again that, even in this case, the
required accuracy (see Refs.~\cite{Martin:2003sg,Martin:2004iv})
renders the numerical computation much longer than the one necessary
to the exploration of the ``full parameter space'' in the standard
inflationary case.

\par

Since we restrict ourselves to the ``fast parameter space'', we have
to fix several degrees of freedom. Firstly, the cosmological
parameters determining the CMB transfer functions should be
chosen. Since the oscillatory features in the multipole moments are
small corrections to the standard inflationary predictions, a
reasonable choice is to fix the (``non primordial'') cosmological
parameters to their best fit values obtained from the vanilla
slow-roll power spectrum, {\ie} from a pure slow-roll model without any
additional feature, in a flat $\Lambda$CDM
universe~\cite{Martin:2003sg,Martin:2004iv,Leach:2003us}. Indeed, it
has been shown in Ref.~\cite{Okamoto:2003wk} that the oscillatory
parameters and cosmological parameters are not very degenerate. This
amounts to take $h=0.734$, $\OmegaB h^2=0.024$, $\OmegaCDM h^2
=0.116$, $\OmegaL=0.740$, $\tau =0.129$, $\zre=14.3 $ and $10^2 \theta
=1.049$, $\theta $ being approximately the ratio of the sound
horizon to the angular diameter distance, see
Ref.~\cite{Lewis:2002ah}.

\par

Secondly, priors on the remaining fast parameters have to be
imposed. For the standard inflationary parameters, $\epsilon_\one$,
$\epsilon_\two $ and the logarithm of the initial amplitude of the
scalar power spectrum at the pivot scale, $\ln \Ps$, wide flat priors
around their standard values have been chosen. Let us now discuss the
three oscillatory parameters. For the phase, the sampling has been
performed on the quantity $\psi $ assuming a flat prior in the
interval $[0,2\pi]$ because there is also a sine function in the term
that we have not written explicitly in Eq.~(\ref{pssrs2}). For the
amplitude of the oscillations, {\ie} $|x| \sigma_\zero$, we have also
imposed a flat prior and have required this parameter to vary in the
range $[0,0.45]$, the upper limit being chosen in order to avoid
negative primordial power spectra.

\begin{figure}
\begin{center}
\includegraphics[height=15.5cm]{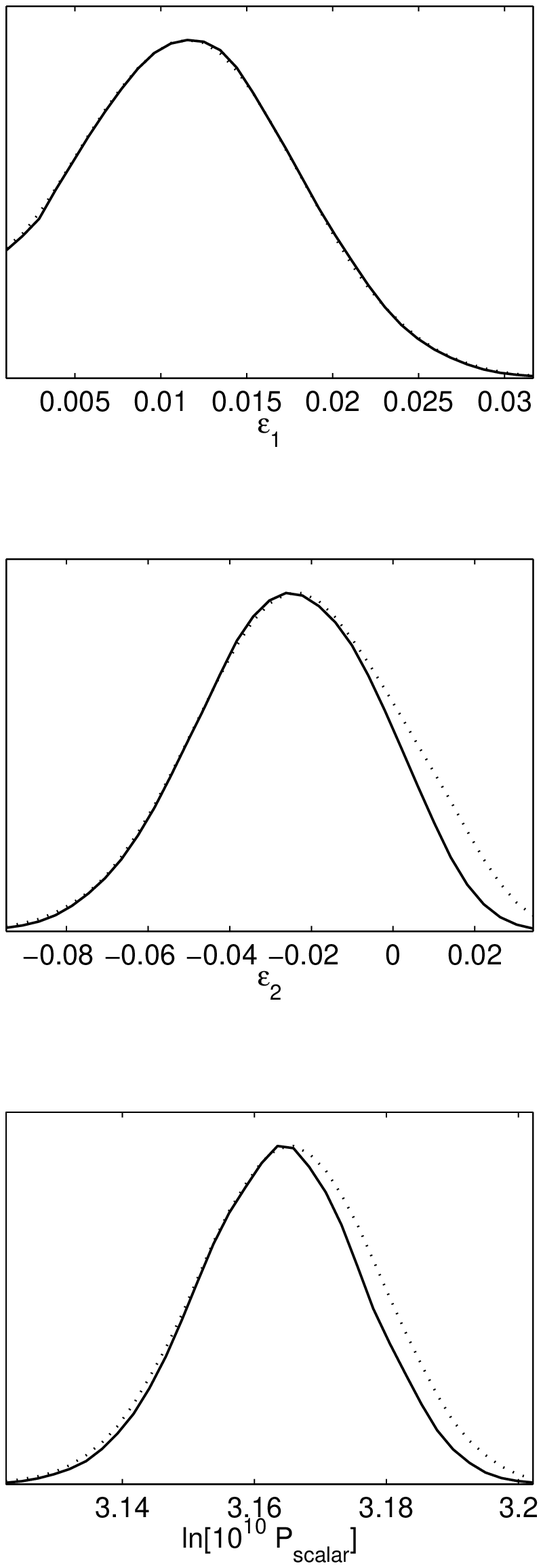}
\hspace{1cm}
\includegraphics[height=15cm,width=6.3cm]{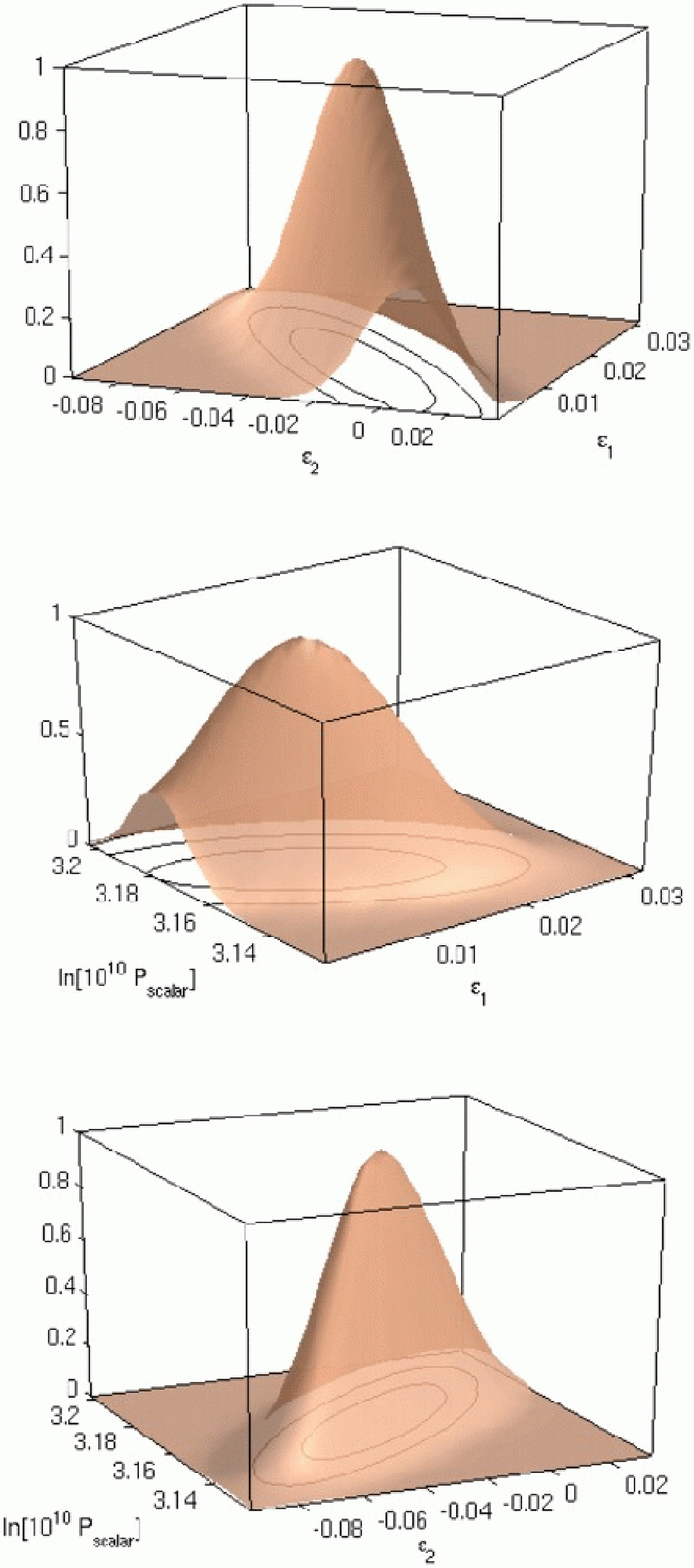}
\end{center}
\caption{$1D$ marginalized probabilities (solid curve) and normalized
mean likelihoods (dashed curve) for the slow-roll parameters
$\epsilon_\one$, $\epsilon_\two$ and the scalar amplitude $\Ps$, in
the case of the reference model. The three-dimensional plots show the
normalized mean likelihoods and the corresponding $1\sigma$ and
$2\sigma$ contours of the $2D$ marginalized probabilities.}
\label{fig:srah}
\end{figure}

\par

The case of the last oscillatory parameter, {\ie} the frequency, should
be treated with great care. The frequency of the superimposed
oscillations is given by $2 \epsilon _\one /\sigma_\zero$ and, hence,
we have chosen to sample the quantity
$\log\left(1/\sigma_\zero\right)$. At this point, it is necessary to
clarify what we exactly mean by ``superimposed oscillations''. Indeed,
if the frequency is very low and is such that less than half of a
period covers the whole observable range of multipoles, then the
resulting effect is just a modification of the overall amplitude of
the power spectra and we don't really have oscillations in the
multipole moments anymore. In this regime, there is a strong
degeneracy between the amplitude and $\Ps$. Clearly, it is not
desirable to reach this extreme case. Another regime corresponds to
the case where the frequency of the superimposed oscillations is of
the same order of magnitude than the frequency of the acoustic
oscillations. In this situation, new peaks would appear in the
$C_{\ell}$'s and this could be problematic since the WMAP data are
well described by the standard inflationary model. As a matter of
fact, the amplitude of this type of oscillations is strongly
constrained, see {\eg} Ref.~\cite{Okamoto:2003wk}. Finally, there is the
case where high frequency oscillations are present. This is the case
we are mostly interested in since it contains the improved fits found
in Refs.~\cite{Martin:2003sg,Martin:2004iv}. Nevertheless, very high
values of the frequency lead to oscillations from multipole to
multipole and any dependency on the frequency is in fact lost. It is
clear that we would also like to avoid this extreme regime. All the
above considerations has led us to restrict ourselves to $\log
\left(1/\sigma_\zero \right)\in [2.3,3.8]$ with a flat prior. This is
equivalent to imposing a Jeffreys-like prior on $\sigma_\zero$.

\par

Our next step has been to generate Markov chains with \COSMOMC for a
standard slow-roll inflationary model without oscillation ({\ie} for our
reference, or vanilla, model) as well as for an oscillatory model
characterized by $\psi$, $|x| \sigma_\zero$ and $\log(1/\sigma_\zero)$
with the above priors, both cases being investigated with the
non-primordial cosmological parameters chosen before. The standard
inflationary model can be used as a reference for the restricted
framework of the ``fast parameter space''. For the two models, the
Markov chains have been started from widespread point in the initial
parameter space and stopped when the posterior distributions no longer
evolve significantly which corresponds to about $250000$ elements.
The generalized Gelman and Rubin $R$-statistics implemented in
\COSMOMC~\cite{Lewis:2002ah,Brooks:1998} is found to be less than
$5\%$ for the slow-roll parameters and equal to approximately $9\%$
for the oscillatory parameters. However, the phase $\psi$ remains
unconstrained. Let us recall that the $R$-statistics gives, for each
parameter, the ratio of the variance of the chain means to the mean of
the chain variances. In other words, it quantifies the errors in the
parameter distributions obtained from the Markov chains exploration.

\section{Results}
\label{sect:constraints}

In Fig.~\ref{fig:srah}, we have plotted the one dimensional
marginalized probabilities and the normalized mean likelihoods for the
slow-roll parameters and the scalar amplitude in the case of the
reference model. Three-dimensional plots of the mean likelihoods,
obtained by averaging with respect to one parameter, are also
represented with the $1\sigma$ and $2\sigma$ contours of the
two-dimensional marginalized probabilities which appear through the
surface. These plots are obviously consistent with the previously
derived constraints~\cite{Leach:2003us}, except that fixing the
cosmological parameters has made the likelihood isocontours slightly
tighter.

\begin{figure}
\begin{center}
\includegraphics[width=12cm,height=16cm]{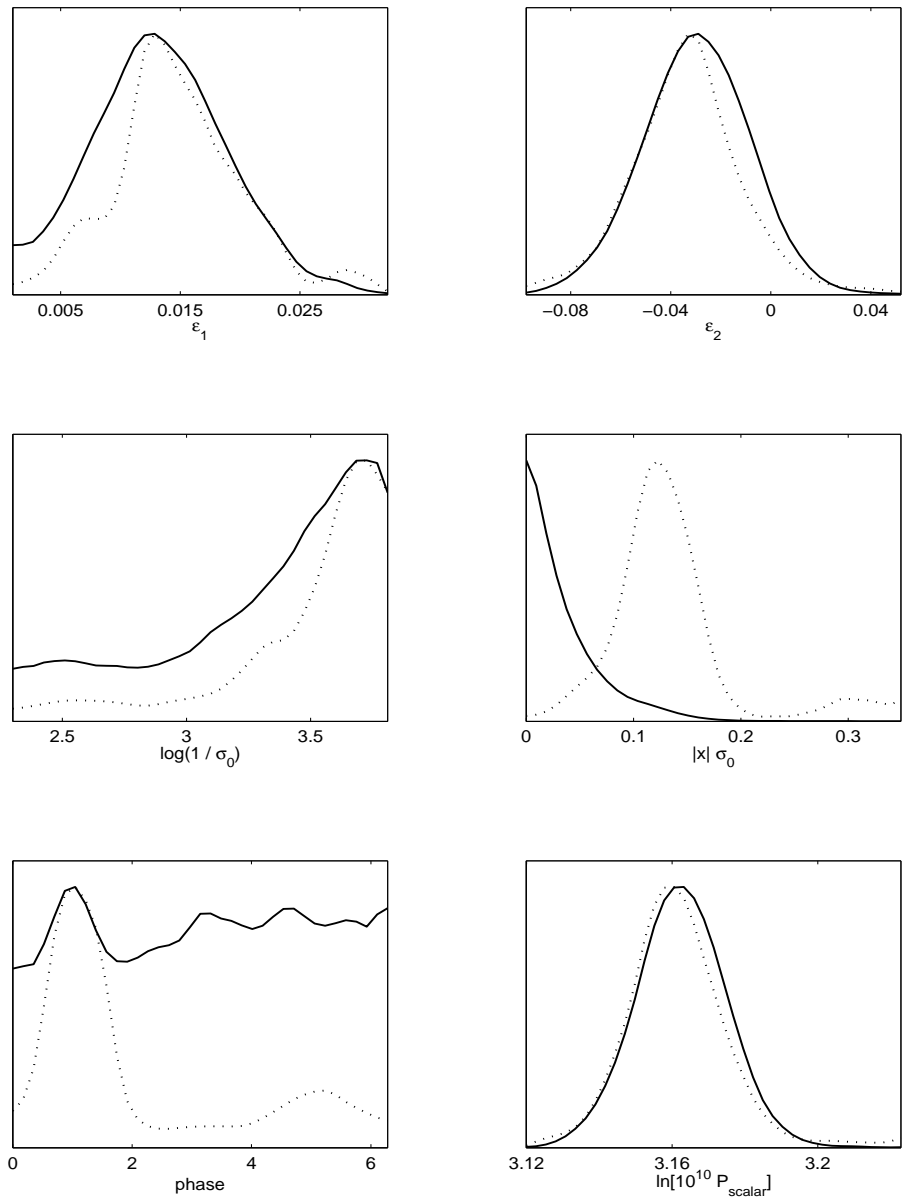}
\end{center}
\caption{$1D$ marginalized probabilities (solid curves) and normalized
mean likelihoods (dashed curves) for the oscillatory model.}
\label{fig:tpl1D}
\end{figure}

\begin{figure}
\begin{center}
\includegraphics[width=6.5cm]{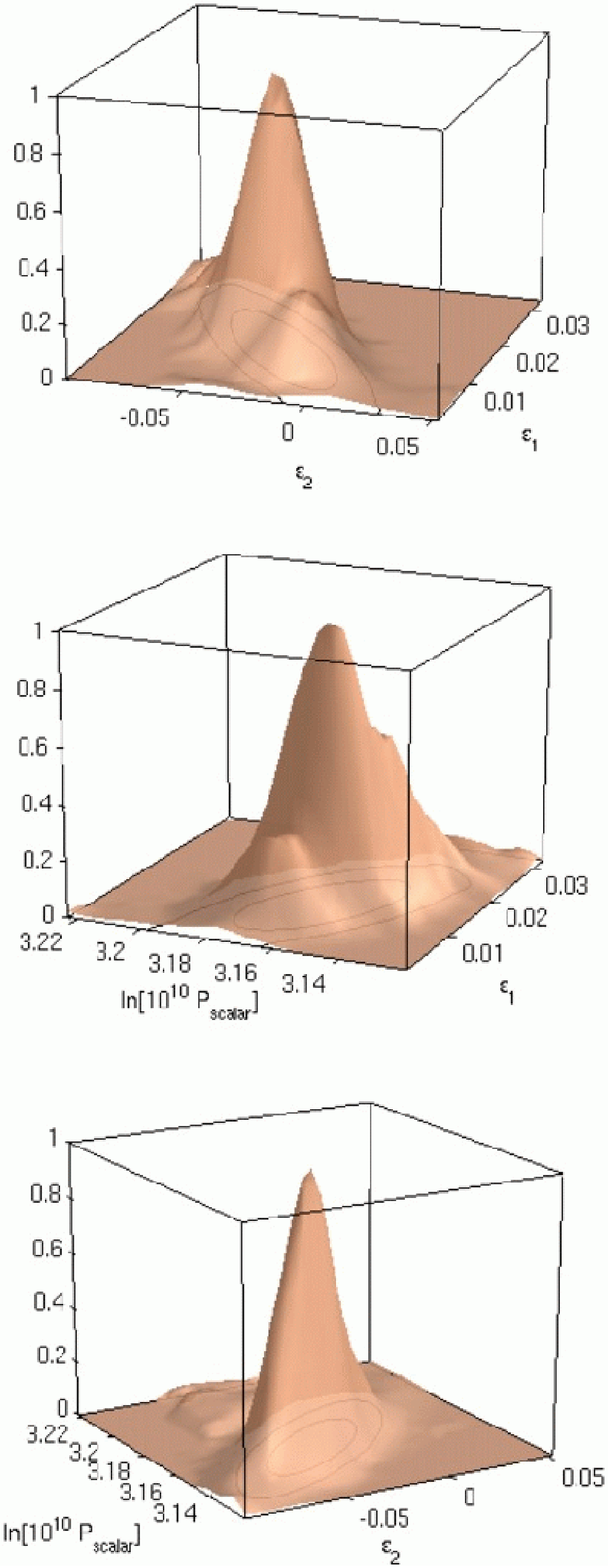}
\hspace{0.5cm}
\includegraphics[width=6.2cm]{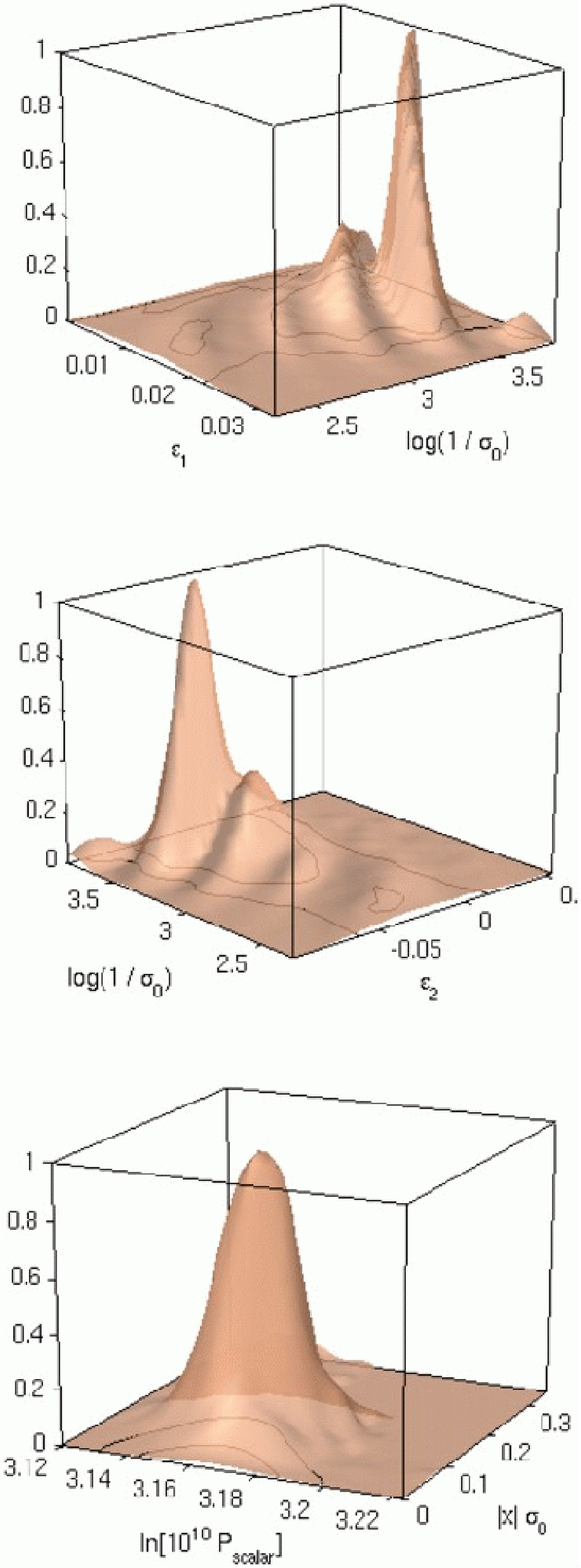}
\end{center}
\caption{$1\sigma$ and $2\sigma$ contours of the $2D$ marginalized
probabilities and normalized mean likelihoods (see also
Fig.~\ref{fig:tpl3Db}).}
\label{fig:tpl3Da}
\end{figure}

\begin{figure}
\begin{center}
\includegraphics[width=6.5cm]{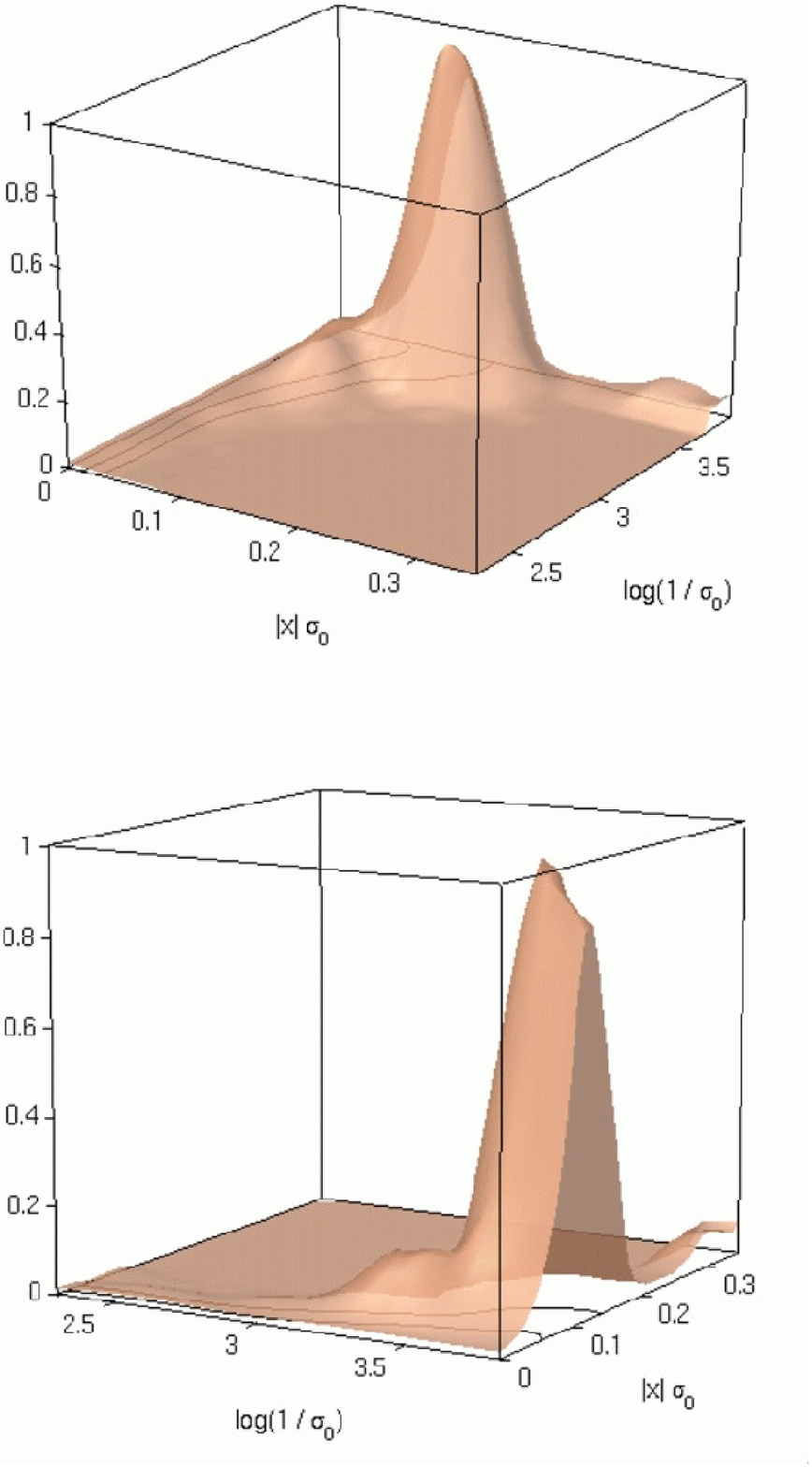}
\hspace{0.5cm}
\includegraphics[width=6.5cm]{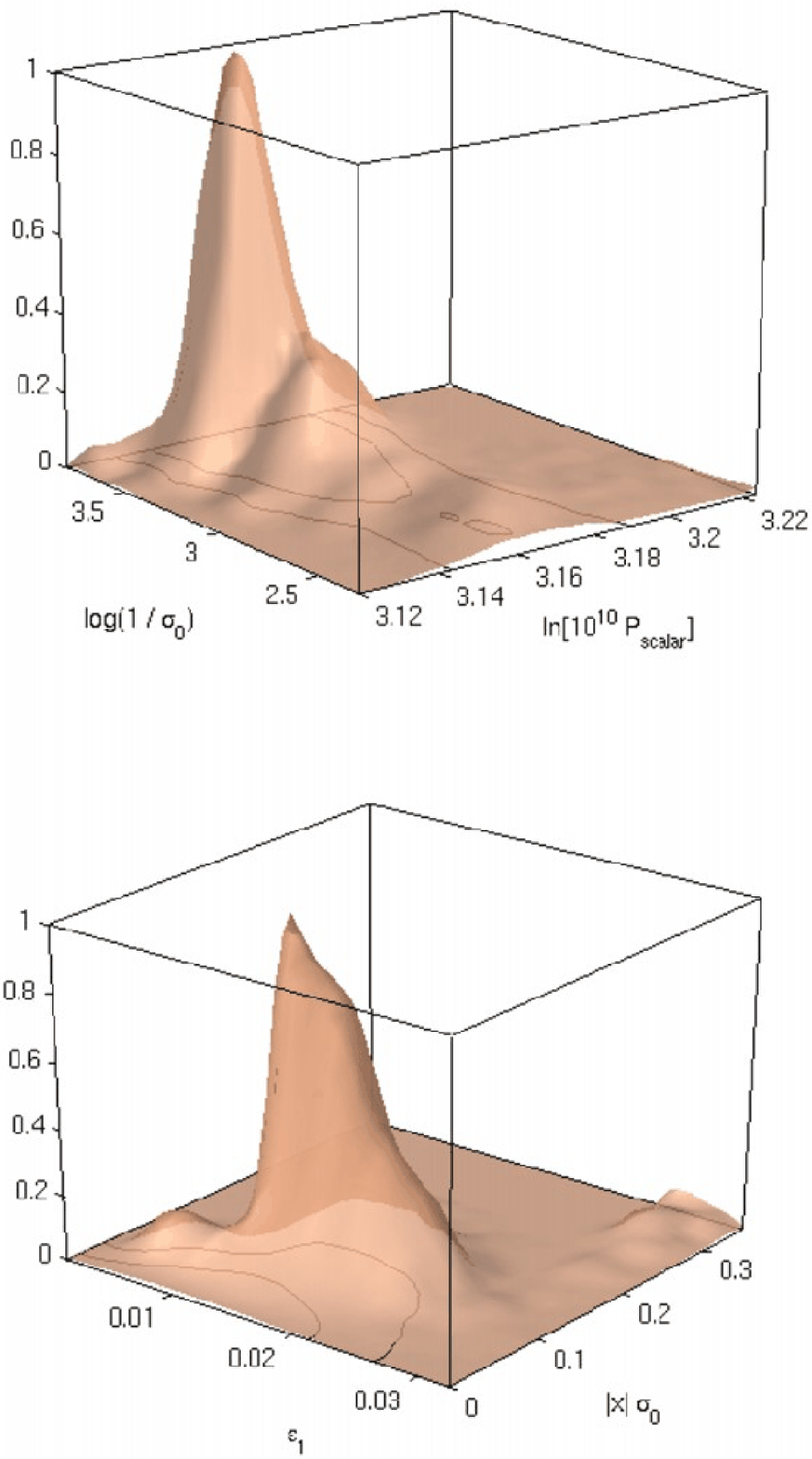}
\end{center}
\caption{$1\sigma$ and $2\sigma$ contours of the $2D$ marginalized
probabilities and normalized mean likelihoods (see also
Fig.~\ref{fig:tpl3Da}). Strong correlations appear between the
amplitude and the frequency of the oscillations (on the left).}
\label{fig:tpl3Db}
\end{figure}

In Fig.~\ref{fig:tpl1D}, we have plotted the $1D$ marginalized
probabilities and the normalized mean likelihoods for the parameters
of the oscillatory model. In Figs.~\ref{fig:tpl3Da}
and~\ref{fig:tpl3Db}, the $1\sigma$ and $2\sigma$ contours of the $2D$
marginalized probabilities and the normalized mean likelihoods are
displayed. They have been plotted in various planes of parameters, the
parameters pairs being chosen to be the most correlated ones.

\par

Let us now discuss the constraints that can be put on the amplitude of
the oscillations, $|x| \sigma_\zero$. As can be seen in
Fig.~\ref{fig:tpl1D}, the vanilla slow-roll model, corresponding to
$|x| \sigma_\zero=0$, remains the most probable one with the currently
available WMAP data. From the probability distribution, one finds that
the $2 \sigma$ marginalized upper limit is given by
\begin{eqnarray}
|x| \sigma_\zero & < 0.11 \, .
\end{eqnarray}
However, the most striking feature of Fig.~\ref{fig:tpl1D} is that the
corresponding mean likelihood does not behave similarly to the
marginalized probability, as it is the case for the other
parameters. Indeed, it exhibits a maximum around $|x| \sigma_\zero
\simeq 0.1$ which is consistent with the best fits found in
Refs.~\cite{Martin:2003sg,Martin:2004iv}. This unusual behavior is
expected if volume effects are present (more precisely, since we
are dealing with quantities defined as integrals over the parameter
space, the volume effects are important when the corresponding kernels
are spread over this space with large values). This highlights the
existence of strong non-linear correlations between $|x| \sigma_\zero$
and the other parameters.

\par

As can be seen in Fig.~\ref{fig:tpl3Db}, strong correlations appear
between the amplitude and the frequency of the oscillations. On the
three-dimensional plot $\left[ |x| \sigma_\zero,\log \left(
1/\sigma_\zero\right) \right]$, one remarks that the mean likelihood
is non-vanishing either for large values of $\log \left(1/\sigma_\zero
\right)$, when $|x| \sigma_\zero \neq 0$, or in a ``long thin tube''
centered around $|x| \sigma_\zero = 0$, the amplitude of which being
much smaller (recall that $\Delta \chi^2 \simeq 10$ between these two
models). This behavior is expected since any frequency is possible
provided its amplitude is very small hence the ``long thin tube''
around $|x| \sigma_\zero=0$. On the other hand, for large values of
$\log \left(1/\sigma_\zero\right)$, one recovers a peak which
corresponds to the oscillations fitting the outliers at relatively
small scales~\cite{Martin:2003sg,Martin:2004iv}. The marginalized
probability is peaked at $\vert x\vert \sigma _\zero =0$ because the
tube has a bigger statistical weight, {\ie} occupies a larger volume in
the parameter space. Note also that this is the very existence of the
likelihood peak at large $\log\left (1/\sigma_\zero \right) $ which
explains the long tail of the $1D$ marginalized probabilities for $|x|
\sigma_\zero$.

On the three-dimensional plot $\left[\epsilon_\one,\log
\left(1/\sigma_\zero \right)\right]$, the wavelets in the mean
likelihood originates from the degeneracy existing between
$\sigma_\zero$ and $\epsilon_\one$, for a fixed value of the frequency
$2 \epsilon_\one/\sigma_\zero$. Furthermore, the correlation between
$\epsilon_\one$ and $\epsilon_\two$~\cite{Leach:2003us} is still
present and, as a consequence, a correlation between $\epsilon_\two$
and $\log\left( 1/\sigma_\zero \right)$ also exists. We also remark
that the correlations appearing when the oscillatory parameters are
taken into account in the analysis seem to slightly tighten the
confidence contours on the standard inflationary parameters and to
cause the appearance of some multimodal features.

\par

Another interesting property concerns the parameter $\log
\left(1/\sigma_\zero \right)$. As can be seen in
Figs.~\ref{fig:tpl1D}, \ref{fig:tpl3Da} and~\ref{fig:tpl3Db}, both the
mean likelihood and the marginalized probability are peaked at a high
value of $\log \left(1/\sigma_\zero \right)$. Quantitatively, a $1
\sigma$ constraint on the energy scale $\Mc$ at which the new physics
shows up can be derived
\begin{eqnarray}
\sigma_\zero &\equiv \dfrac{H}{\Mc} < 6.6 \times 10^{-4}\, .
\end{eqnarray}
Let us stress that this constraint is valid only if the amplitude and
the frequency of the superimposed oscillations are considered as
independent parameters. Indeed, it mainly comes from the region where
the likelihood is strongly peaked and, as already noticed, this region
does not exist if the parameter $x$ is chosen such that $x=1$. In this
case, the likelihood is rather flat and, as a consequence, one would
obtain a weaker contraint on the ratio $H/\Mc$. 

\par

Finally, we have found that the phase $\psi$ is not constrained as is
clear from Fig.~\ref{fig:tpl1D}.

\par

To end this section, one can ask how well the oscillatory model fits
the data \emph{on average}, compared to the standard slow-roll
one. This can be estimated by comparing the mean likelihood over all
the parameters in each of these models~\cite{Lewis:2002ah}. For the
reference model one gets $\ln \Lcb_\sr \simeq -715.1$ while the
oscillatory model has $\ln \Lcb_\wig \simeq -712.9$, hence a ratio of
$\exp\left(715.1-712.9\right)\simeq 9$ in favor of the oscillatory
model. This shows that including the oscillatory parameters permits to
improve the goodness of fit on average~\cite{Lewis:2002ah}.

\section{Discussion and Conclusion}
\label{sect:conc}

In this paper, we have explored, by means of Monte Carlo methods, the
fast parameter space of an inflationary model with superimposed
oscillations. This restricted framework is, at the time of writing,
the only way to derive constraints in a reasonable computational time
if the region where high frequency oscillations that better fit the
CMB anisotropies outliers at relatively small
scales~\cite{Martin:2003sg,Martin:2004iv} is included. Among the main
results derived in the present article are two constraints, one on the
amplitude, $\vert x\vert \sigma _\zero <0.11 $, and the other of the
energy scale $\Mc$, $H/\Mc <6.6 \times 10^{-4}$.

\par

The overall situation is quite interesting: on one hand, the fact that
the marginalized probability is peaked at a value corresponding to a
vanishing amplitude shows that the most probable model remains the
standard slow-roll one for which $\vert x\vert \sigma _\zero =0$. On
the other hand, this distribution exhibits a long tail in the regions
corresponding to non-vanishing amplitudes of the oscillations,
precisely where the mean likelihood function is peaked, {\ie} around
$\vert x\vert \sigma _\zero \simeq 0.1$. Moreover, the ratio of the
vanilla slow-roll model to the oscillatory model total mean
likelihoods is about $9$, in favor of the oscillatory model. This
shows that, on average, the oscillatory model fits better the first
year WMAP data.

\par

The interpretation of the situation described above is as follows. The
marginalized probabilities are quantities which are especially
sensitive to ``volume effects'', {\ie} to the shape of the likelihood
in regions of the parameter space where this one is significant. On
the other hand, the mean likelihood is sensitive to the absolute value
of the likelihood regardless to its occupied ``volume''. Therefore, if
in the parameter space there are regions which, at the same time,
correspond to good fits and occupy a quite confined volume, then these
regions will appear particularly significant from the mean likelihood
point of view but will be ``diluted'' and, hence, will appear less
significant from the marginalized probabilities point of
view. Ultimately, the marginalized probability is the relevant
quantity, that is to say is the quantity which should be used in order
to conclude about the statistical meaning of a given
model~\cite{Aitkin:1991}. It will be interesting to study how the
situation evolves when better data become available, in particular it
will be interesting to see if the models with oscillations can occupy
a volume which makes them probable from the marginalized probabilities
point of view. This is one of the reasons why the next WMAP data
release will be of great interest.

\acknowledgments

We wish to thank Samuel Leach, Antony Lewis and Eugene Lim for
enlightening discussions and useful correspondences. We would like to
thank the IDRIS for providing us numerical resources.

\section*{References}

\bibliography{bibtpl}

\end{document}